\documentclass{aa}  
\usepackage{graphicx}
\usepackage{graphics}
\usepackage{amssymb}
\usepackage{txfonts}
\usepackage{mathrsfs}
\usepackage{array}
\usepackage{ulem}
\usepackage{longtable,lscape}
\usepackage{array} 
\usepackage{tabularx}
\usepackage{amsmath}
\usepackage{natbib}
\bibpunct{(}{)}{;}{a}{}{,}   % to follow the A&A style
\begin{document}
\title{An Intriguing Globular Cluster in the Galactic Bulge from the VVV Survey
\thanks{Based  on  observations  taken within  the  ESO  VISTA  Public Survey VVV, Programme ID 198.B-2004}}
%   \subtitle{}

\author{
D. Minniti \inst{1,2}
\and T.~Palma \inst{3} 
\and D.~Camargo \inst{4}  
\and M.~Chijani-Saballa \inst{1} 
\and J.~Alonso-Garc\'{i}a \inst{5, 6} 
\and J.~J.~Clari\'a \inst{3} 
\and B.~Dias \inst{7} 
\and M.~G\'omez \inst{1}  
\and J. B.~Pullen \inst{1}  
\and R.~K.~Saito \inst{8} 
}

\institute
{
$^1$ Departamento de Ciencias F\'{\i}sicas, Facultad de Ciencias Exactas, Universidad Andres Bello, Fern\'andez Concha 700, Las Condes, Santiago, Chile \\
$^2$ Vatican Observatory, Vatican City State, V-00120, Italy\\
$^3$ Observatorio Astron\'omico, Universidad Nacional de C\'ordoba, Laprida 854, 5000 C\'ordoba, Argentina\\
%$^4$ Colegio Militar de Porto Alegre, Ministerio da Defesa, Av. Jos\'e Bonif\'acio 363, Porto Alegre 90040-130, RS, Brazil\\
$^4$ Col\'egio Militar de Porto Alegre, Minist\'{e}rio da Defesa, Av. Jos\'{e} Bonif\'{a}cio 363, Porto Alegre, 90040-130, RS, Brazil\\
$^5$ Centro de Astronom\'{i}a (CITEVA), Universidad de Antofagasta, Av. Angamos 601, Antofagasta, Chile\\
$^6$ Millennium Institute of Astrophysics, Nuncio Monse\~nor Sotero Sanz 100, Of. 104, Providencia, Santiago, Chile\\
$^7$ Instituto de Alta Investigaci\'on, Sede Iquique, Universidad de Tarapac\'a, Av. Luis Emilio Recabarren 2477, Iquique, Chile\\
$^8$ Departamento de F\'{i}sica, Universidade Federal de Santa Catarina, Trindade 88040-900, Florian\'opolis, SC, Brazil
}

%\offprints{to: email@email} 
  
 \date{Received ; Accepted }


\abstract
%  context (optional) 
{Globular clusters (GCs) are the oldest objects known in the Milky Way so each discovery of a new GC is astrophysically important. 
In the inner Galactic bulge regions these objects are difficult to find due to extreme crowding and extinction. 
However, recent near-IR Surveys have discovered a number of new bulge GC candidates that need to be further investigated.
}
%  aims
{Our main objective is to use public data from the Gaia Mission, the VISTA Variables in the Via Lactea Survey (VVV),  the Two Micron All Sky Survey (2MASS), and the Wide-field Infrared Survey Explorer (WISE)  in order to measure the physical parameters of Minni\,48, a new candidate globular star cluster located in the inner bulge of the Milky Way at $l=359.35$ deg, $b=2.79$ deg.
The specific goals are to measure its main astrophysical parameters like size, proper motions, metallicity, reddening and extinction, distance, total luminosity and age.
}
% method
{Even though there is a bright foreground star contaminating the field, this cluster appears quite bright in near- and mid-IR images. 
The size of Minni\,48 is derived from the cluster radial density profile, while its reddening and extinction are estimated from optical and near-IR reddening maps. We obtain %deep 
statistically decontaminated optical and near-IR colour-magnitude diagrams (CMDs) for this cluster. Mean cluster proper motions are measured from Gaia  data. The heliocentric cluster distance is determined from both the red clump (RC) and the red giant branch (RGB) tip magnitudes in the near-IR CMD, while the cluster metallicity is estimated from the RGB slope and the fit to theoretical stellar isochrones.}
% results 
{The size of this GC  is found to be $r = 6’ \pm 1' $, while 
the reddening and extinction values are $E(J-K_s)=0.60 \pm 0.05$ mag, $A_G=3.23 \pm 0.10$ mag, $A_{Ks}=0.45 \pm 0.05$ mag. 
The resulting mean cluster proper motions are $\mu_{\alpha}= -3.5 \pm 0.5 $ mas/yr, $\mu_{\delta}=-6.0 \pm 0.5 $  mas/yr.
We also study the RR Lyrae stars recognized in the field and we argue that they are not members of this GC. 
The magnitude of the RC in the near-IR CMD yields an accurate distance modulus estimate of 
$(m-M)_0=14.61$ mag, equivalent to a distance $D=8.4 \pm 1.0$ kpc.
Such distance is consistent with the optical distance estimate: 
$(m-M)_0=14.67$ mag, $D=8.6 \pm 1.0$ kpc, and also with the distance estimated using the tip of the RGB: 
$(m-M)_0=14.45$ mag, $D=7.8 \pm 1.0$ kpc.
We also derive a cluster metallicity of $[Fe/H]=-0.20 \pm 0.30$ dex.
Adopting these values of metallicity and distance, a good fit to the PARSEC stellar isochrones is obtained in all CMDs using $Age = 10 \pm 2$ Gyr.
The total absolute magnitude 
of this GC is estimated to be 
$M_{Ks}= -9.04 \pm 0.66$ mag.
}
% conclusion
{Based on its position, kinematics, metallicity and age, we conclude that Minni\,48 is a genuine GC, similar to other well known metal-rich bulge GCs.
It is located at a projected  Galactocentric angular distance of 2.9 deg, equivalent to 0.4 kpc, situating this cluster as one of the closest GCs to the Galactic centre currently known.
}

\authorrunning{Minniti  \& et al.}
\titlerunning{A New VVV Globular Cluster in the Galactic Bulge} 
\maketitle

\section{Introduction}

The inner Galactic bulge globular clusters (GCs), in particular, are very difficult to find and to characterize at optical wavelengths due to heavy interstellar extinction and severe stellar crowding. Recent near-IR Surveys (the Two Micron All Sky Survey (2MASS) - Skrutskie et al. 2006, and the VISTA Variables in the Via Lactea Survey (VVV - Minniti et al. 2010)), and mid-IR Surveys (the Spitzer Galactic Legacy Infrared Mid-Plane Survey Extraordinaire (GLIMPSE - Benjamin et al. 2003), and the Wide-field Infrared Survey Explorer (WISE - Wright et al. 2010)) have meant a real revolution for the study of GCs in the Milky Way (MW) bulge. 

On one hand, the deep near and mid-IR images led  to the discovery of dozens of new GC candidates in the bulge that had hitherto remained hidden due to the high crowding and severe differential reddening in the inner regions of the Galaxy. For example, Camargo et al. (2015, 2016) developed a star cluster Survey using WISE images, which allowed for the discovery of eight new bulge GCs (Camargo 2018, Camargo \& Minniti 2019).

On the other hand, the near and mid-IR photometry and the multi-epoch observations of variable stars have permitted the accurate measurement of the physical parameters (sizes, reddenings, distances, ages, luminosities, etc.) for several known GCs that have been so far studied (e.g., 
Longmore et al. 2011,
Minniti et al. 2011,
Moni Bidin et al. 2011,
Ortolani et al. 2012,
Alonso-García et al. 2015,
Borissova et al. 2014, 
Minniti et al. 2017a, 2017b, 
Piatti 2018,
Ryu \& Lee 2018, 
Palma et al. 2019).
These studies have contributed to a deeper understanding of the formation and evolution of the MW GC system.

At the distance of the bulge, the GLIMPSE and WISE mid-IR Surveys are sensitive to the clusters' bright red giants (RGs).
The 2MASS catalogue allows to define the cluster upper red giant branch (RGB), used to measure the RGB tip and the RGB slope.
The VVV photometry samples deeper into the RGB, making it possible to measure the horizontal branch (HB) shape and the red clump (RC) magnitude. 
These near and mid-IR Surveys, in combination with the optical photometry and proper motions (PMs) from the Gaia mission (Brown et al., Gaia Collaboration 2018), can be used to measure the cluster physical parameters,  including position, PMs, size, metallicity, luminosity, reddening, extinction and also age using isochrones fitted to the observed cluster sequence.

In this paper we concentrate on a fairly rare object, the candidate GC Minni\,48. Using the VVV Survey, this candidate GC was discovered as a concentration of RR Lyrae and RGs, suggesting a metal-poor GC (Minniti et al. 2018). However, later examination of the mid-IR GLIMPSE and WISE images shows a cluster that appears quite bright at these wavelengths, 
suggesting a metal-rich GC. 
Interestingly, the Gaia EDR3 CMD shows that the upper giant branch in the cluster region is well split into a bluer and a redder RGB, as shown in Figure 1
(the Gaia DR2 CMD shows a similar upper RGB split).

This cluster is located along the Galactic minor axis (l=-0.65 deg) at low latitude (b=+2.79 deg). This is one of the closest clusters to the Galactic centre, with a projected Galactocentric angular distance r=2.9 deg. A major complication for its study is the high field stellar density wherein it is projected, which is even higher than that of Baade's window at $b=-4$ deg, a window of low extinction also located along the minor axis but on the other side of the Galactic plane, where the classical GCs NGC 6522 and NGC 6528 are found. 

Another obstacle for the study of this object is the presence of a very bright foreground stellar source (IRAS 17302-2800, a.k.a. 2MASS 17331634-2800145), projected near its centre.
With magnitudes $K_s=3.33$, $G=10.57$, and colour indexes $(J-K_s)=1.81$, $(BP-RP)=5.60$, this star  is badly saturated in optical and IR images, affecting the photometry of the surrounding region. In fact, we argue that Minni\,48  is an ordinary GC that has been  previously missed due precisely to the presence of this bright source.
On the other hand, note that even in the presence of a bright foreground star, it is possible to recover the underlying cluster, as shown for example by the discovery of one of the first Gaia clusters (Koposov et al. 2017).

\section{The Optical and IR Datasets}
\label{sec:area_obs}

Given that the stellar density is so high in the cluster region, we must make sure of the measured location of the RGB and RC, which determine important cluster parameters such as reddening, distance and metallicity. For this reason, we use all available optical and near-IR  data. Fortunately, we are able to come to a consistent solution across all wavelengths.

The examination of the GC appearance in the sky reveals a clear excess of stars with respect to the background. This excess is more conspicuous at the RGB tip in the 2MASS CMDs and at the RC regions of the VVV and DECAPS CMDs,
although at the shorter wavelengths it is more difficult to appreciate because it is eclipsed by the bright star (Figure 2).
Such excess is prominent when we restrict the comparison to stars kinematically selected using Gaia EDR3 PMs.
A brief description of the available datasets along with the relevant references is presented below.

\subsection{ 2MASS photometry for bright stars}
 
 The 2MASS is an all sky Survey in the $JHK_s$ near-IR bands (Skrutskie et al. 2006, Cutri et al. 2003). We retrieved the 2MASS data from the VizieR Online Data Catalogue.
Since the brightest stars in Minni\,48 ($K_s<11$ mag) appear to be saturated in the VVV near-IR images (Saito et al. 2012), we use 2MASS data in order to build the complete CMD. 

The first selection applied here culls stars with large photometric errors: $\sigma_J > 0.3$ mag, and  $\sigma_{Ks} > 0.3$ mag. 

The 2MASS photometry is also linearly related to the VVV photometry (for a detailed description see Mauro et al. 2013, Soto et al. 2013 and Gonzalez-Fernandez et al. 2018).

\subsection{Deep VVV Survey observations}

The VVV Survey and its extension, the VVVX Survey, are large ESO public near-IR surveys of the southern MW plane (Minniti et al. 2010; Minniti 2018).  Data reduction was carried out at the Cambridge Astronomical Survey Unit (CASU - Irwin et al. 2004, Lewis et al. 2010). The point spread function (PSF) photometry was extracted using the pipeline described in Alonso-García et al. (2018). The resulting catalogues are in the VISTA photometric system, which is different but related to the 2MASS photometric system (Mauro et al. 2013; Soto et al. 2013; González- Fernández et al. 2018)

Minni\,48 is located in the VVV tile b361, a very reddened and crowded bulge field. No photometric or colour cuts were applied to its VVV PSF photometry. As in our analysis the catalogues from 2MASS and VVV were treated independently, we kept the VVV catalogue for Minni\,48 in its original VISTA photometric system. We also note that the present VVV calibration should be trusted to $\Delta JKs < 0.05$ mag, because some field zero-point variations have been reported in the VVV photometry (Hajdu et al. 2019), although in the present case such difference could also be partly due to the influence of the bright foreground source IRAS 17302-2800 that affects the background in the VVV images.

\subsection{GLIMPSE and WISE Mid-IR photometry} 

The GLIMPSE is a Survey of the Galactic plane in four mid-IR bands centred at 3.6, 4.5, 5.8 and 8.0 microns (Benjamin et al. 2003, Churchwell et al. 2009). 
The WISE is an all sky Survey (Wright et al. 2010) that uses four mid-IR bands $W1, W2, W3, W4$, centred at $3.3, 4.7, 12$ and $24$ microns, respectively, with lower spatial resolution ($6"-10"$).
Visual inspection of the WISE images (Figure 3) reveals that Minni\,48 is prominent and bright in the mid-IR, exhibiting a compact core of bright stars.
This is confirmed also by the  GLIMPSE image data that has higher resolution ($1"-2"$). 

We retrieved from the VizieR Online Data Catalogue the WISE data for the four mid-IR bands $W1, W2, W3, W4$. 
Figure 4 shows the spatial distribution of WISE sources in Galactic coordinates and the corresponding WISE mid-IR CMD. 
This WISE CMD exhibits the presence of very red stars, generally indicative of evolved metal-rich stars.
We find that there is a clear concentration of many RGs in the cluster core (within 3’ of the cluster centre).
There are ten very red stars with $W1-W2>0.1$ mag in a 10’ field centred on the cluster, with
five out of these red stars being within 3’ of the centre.
The brightest star with $W1=2.1$ mag is the foreground source IRAS 17302-2800. 

\subsection{Gaia optical photometry and proper motions}

We use Gaia DR2 data (Gaia Collaboration 2018) and EDR3 data (Brown et al. 2021) retrieved from the VizieR Online Data Catalogue.
The first selection excludes all nearby foreground stars with $plx > 0.5$ mas (i.e., $D<3$ kpc, Mignard et al. 2018, Bailer Jones et al. 2018).
No photometric colour or magnitude cuts were applied to the Gaia data.

\subsection{DECAPS Optical photometry}

The DECAPS is an optical Survey of the Southern Galactic plane and bulge fields with $-5<b<5$ deg (Schlafly et al. 2017).
We use the DECAPS fluxes in $griYz$  converted to magnitudes, retrieved from their webpage (Schlafly et al. 2017).
This optical PSF photometry is complementary, because it provides different colours, and it is deeper than the Gaia photometry in this field ($r_{lim} \sim 24$ mag vs $G_{lim} \sim 21$ mag, respectively).

\section{Globular Cluster Physical Parameters}
\label{sec:parameters}

There are some major problems for the study of this GC candidate. First and foremost, there is the issue of high field stellar density.
This is a common problem for the study of GCs in the bulge, but it is enhanced in the inner Galactic field studied here.
For example, using the deep VVV photometry, we count that there are 66\% more RC giants per sqdeg in this field than at Baade's window, where the well known metal-rich GCs NGC 6522 and NGC 6528 are located. 
This high field contamination (both foreground and background) affects the measurement of the cluster size and total luminosity 
and is reflected in the large errors that we estimate for these two parameters.

Other serious problems are the  high differential reddening and extinction. The shape of the extinction law is known to vary in the inner bulge (e.g., Majaess et al. 2016, Nataf et al. 2016, Alonso-Garc\'ia et al. 2017).
This makes it difficult to determine the GC distance and increases the systematic errors. However, different methods (RC in the optical and near-IR and the RGB tip) can be used in this study to check consistency.
The uncertainty in the extinction also affects age determination using isochrones, thus yielding large errors. Fortunately, the availability of multi-wavelength photometry allows us to reach a consistent solution, as discussed below.

\subsection{Colour-Magnitude Diagrams}

As mentioned before, the very high field contamination (foreground and background) in the cluster field is a serious problem for the determination of the cluster parameters.
Giving the variety of datasets for this cluster (VVV, 2MASS, Gaia, DECAPS), we deemed appropriate to make two independent decontamination experiments.
The cluster parameters can then be better determined using the decontaminated CMDs.

{
Thus, aiming to uncover the intrinsic CMD morphology of Minni\,48 from the bulge field stars, we apply a field-star decontamination procedure. The algorithm divides the CMDs of the cluster region and of a comparison field into a 3D grid (a magnitude and two colours) measuring the density of stars with compatible magnitude and colours in the correspondent cells. Then, the algorithm removes the expected number of field stars from each cell of the cluster CMD (Bonatto \& Bica 2007; Bica et al. 2008; Camargo et al. 2009; Camargo et al. 2016). In this work we apply a version of the algorithm to the Gaia photometry.
Additionally, we used a similar procedure developed by Palma et al. (2016). These decontamination procedures differ in the comparison field selection, the first used a ring of large area centred in the cluster coordinates while the second used equal area background fields with similar reddening $\sim 10'$ away from the cluster for this purpose.
}

Figure 5 shows the
Gaia EDR3 optical CMDs ($G$ vs $BP-RP$) for the GC region (left panel), a background region covering a similar area on the sky as the cluster region (middle panel), and the statistically decontaminated CMD (right panel), following 
the procedures of { Camargo et al. (2009)} and Palma et al. (2016).
{
It is expected that the observed CMD of a GC embedded in the bulge dense field  that remained hidden until recently shows some similarity to the near-field CMD.
Sometimes it is hard to clearly perceive the differences between these CMDs by eye, but the decontamination algorithm works on a statistical basis and is able to uncover the cluster intrinsic CMD by comparing colours and magnitudes of the two stellar contents.
}
 
The decontaminated CMD on the right panel of Figure 5 shows  that only the RGB (more metal-rich) survives the decontamination procedure.
The bright portion of this optical CMD is easily decontaminated because there is a clear overdensity of luminous cluster stars. 
The faintest portion, however, is much more difficult to decontaminate because the bright star opaques the fainter ones, and as a result, there is not a clear overdensity with respect to the background. { 
In addition, the accuracy of the photometry may be affected by limitations on the data processing and variations of the local background, in the sense that the BP and RP fluxes may include contributions of the nearby bright sources. The highest impact occurs for crowded environments like the inner regions of globular clusters and the Galactic Bulge. The faint sources are most strongly affected, especially those near bright stars (Arenou et al. 2018; Evans et al. 2018). On the other hand, obviously,  Gaia still provides a substantial gain in relation to 
 previous large sky surveys.}

Figure 6 shows the
2MASS near-IR CMDs ($K_s$ vs $J-K_s$) for the GC region (left panel), a background region having a similar area on the sky as the GC region (middle panel), and the statistically decontaminated CMD following 
Bica et al. (2008) and Camargo et al. (2009) (right panel). 
The decontaminated CMD on the right  panel clearly illustrates the metal-rich RGB and the RGB tip located at $K_s=7.75 \pm 0.10$ mag. 

Figure 7 presents the
deep VVV near-IR CMDs ($K_s$ vs $J-K_s$) for the GC region (left panel), a background region covering a similar area on the sky as the cluster region (middle panel), and the statistically decontaminated CMD following 
the steps described in detail by Palma et al. (2016) (right panel). 
The decontaminated CMD in the right panel clearly shows the  RC of this GC at $K_s=13.45$ mag. 

This feature is important because RC stars are well known distance indicators (e.g., Girardi 2016), and will be used here to determine the GC distance, adopting 
$Ks_{RC}= -1.606 \pm 0.009$ mag and 
$G_{RC}= 0.383 \pm 0.009$ mag  (Ruiz-Dern et al. 2018).

All these CMDs (Figures 5, 6, 7) show that the decontaminated sequence for the cluster is much tighter than that of the background. These datasets have been treated independently, and they all confirm the presence of a real star cluster.

In order to complement the Gaia optical CMDs, Figure 8 shows the
optical CMDs for the GC region from the DECAPS Survey (Schlafly et al. 2017). 
These CMDs are zoomed in the RGB region. They all clearly show the GC RC at $r = 18.65$ and $i = 17.50$ mag.
We can also appreciate the RGB bump in these diagrams.

We note that in the presence of differential reddening, if the statistical decontamination procedure leaves abundant field contamination, the mean reddening and metallicity values may be biased.
There is another independent way to decontaminate the cluster CMD, and that is using the exquisite  astrometry provided by the Gaia mission. 
In order to do this, we first select all stars within 3' from the cluster centre with parallaxes smaller than 0.5 mas. 
This selection eliminates nearby stars belonging to the Galactic disk.
After identifying the PM peak 
due to the GC members in the VPM diagram (Figure 9), we select GC members within 2 mas/yr, centred on these PMs.
{
The mean cluster PMs measured from Gaia EDR3 are $\mu_{\alpha}= -3.5 \pm 0.5 $ mas/yr, $\mu_{\delta}=-6.0 \pm 0.5 $  mas/yr.}
These PM-selected members were then matched with the VVV near-IR photometry and the resulting CMDs are shown in Figure 11. 
These CMDs are similar to those obtained using the statistical background decontamination procedure (Figures 5, 6, 7). 
They almost reach the main sequence turn-off of the cluster, clearly showing a tight RGB and a prominent RC.
Figure 12 shows the luminosity functions in the G and Ks-bands, where the RC is clearly seen.
This PM decontamination procedure was performed first with Gaia DR2 data and repeated again using Gaia EDR3 data, thus yielding consistent results.

{
Figure 10 shows the Gaia EDR3 PM distribution for the Minni\,48 central region and for a ring in the comparison field. Minni\,48 presents a PM distribution consistent with that of a typical GC located in the Galactic bulge (Camargo 2018, Camargo \& Minniti 2019).  Figure 10 shows a more compact PM distribution for stars within the cluster area with respect to those in the comparison field, suggesting a  cohesive structure. As most bulge GCs, Minni\,48 follows the local bulge stellar content (Geisler et al. 2021, Minniti et al. 2021a,b).}

\subsection{Measurement of cluster reddening and absorption}

It is not straightforward to find all the parameters that fit simultaneously all CMDs after the  CMD decontamination is made.
This difficulty is caused in part by the uncertainties in the cluster reddening and extinction. 
These parameters are estimated from existing optical and near-IR reddening maps. The most recent reddening and extinction measurements for the cluster field are in reasonable agreement. The VVV extinction maps (Gonzalez et al. 2011, Surot et al. 2020) yield a colour-excess $E(J-K_s) = 0.75$ mag for this field. The associated near-IR extinctions are $A_K=0.52$ and $0.40$ mag for the Cardelli et al. (1989) and Nishiyama et al. (2009) extinction laws, respectively.
On the other hand, the maps of Schlafly et al. (2011) give a reddening $E(J-K_s) = 0.61$ mag and an extinction $A_K=0.45$ mag.
We finally adopt the values $E(J-K_s) = 0.61 \pm 0.03$ mag and $A_K = 0.45 \pm 0.05$ from Schlafly et al. (2011),  intermediate between the two determinations of Gonzalez et al. (2011) for this field.

As a caveat, interstellar reddening is patchy and the choice of a mean reddening is justified if there are no significant reddening variations.
However, this field does not seem particularly complicated in this regard, because the 2D maps of Surot et al. (2020) show that there are no strong indications of severe differential reddening
($\Delta (J-Ks)<0.10$ mag), and also because the 3D maps of Chen et al. (2013) show that most of the clouds that cause the reddening along this line of sight appear to be located in the foreground.

An external control on the adopted cluster reddening value is made using the RR Lyrae stars  in the field that have a mean colour  index $\langle J-K_s \rangle_{RRL}=0.8$ mag, yielding $E(J-K_s)=0.65$ mag, which is fully consistent with the value adopted here from the published extinction maps.
The corresponding extinction $A_{Ks}=0.45 \pm 0.05$ mag is equivalent to $A_G = 0.79 \times A_V = 3.23 \pm 0.10$ mag in the Gaia passbands.

\subsection{Distance determination}

With the reddening and extinction values at hand, we use the position of the RC in the optical and near-IR CMDs to measure the cluster distance. To do this, we adopt the intrinsic RC magnitude 

$Ks_{RC}= -1.606 \pm 0.009$ mag and 
$G_{RC}= 0.383 \pm 0.009$ mag 
from Ruiz-Dern et al. (2018).
The observed optical and near-IR  luminosity functions are shown in Figure 11.
 We use the RC position in the near-IR that is less sensitive to reddening variations, after proper decontamination using both the CMD and the Gaia PMs. The mean RC magnitudes for both decontaminated samples shows good agreement, 
so we adopt $Ks_{RC}=13.45 \pm 0.10$ mag. 

Using the RC measured from the optical CMD of the Gaia EDR3 data, the mean RC magnitude is $G=18.40 \pm 0.10$ mag. Schlafly et al. (2011) give an optical extinction of $A_V=4.08$ mag. The extinction adopted here is $A_G=0.79*Av=3.23$ mag, yielding a dereddened RC optical magnitude of $G_0=15.17$.
Therefore, the resulting cluster distance modulus from the optical photometry is $m-M= 14.67 \pm 0.15$ mag, equivalent to a distance $D=8.6 \pm 1.0$ kpc, 
that is in excellent agreement with that measured using the VVV near-IR photometry $m-M= 13.45+1.61-0.45 =14.61 \pm 0.15$ mag.

The brightest cluster giants have $K_s=7.70 \pm 0.2$ mag and $(J-K_s)=2.0 \pm 0.1$ mag, which we take as the magnitude and colour of the RGB tip. 
Adopting the calibration of Serenelli et al. (2017) for near-IR passbands, the resulting distance modulus is m-$M=14.45 \pm 0.2$ mag, equivalent to $D=7.8$ kpc.

This value is somewhat shorter than the distance measured from the RC stars in the optical and near-IR CMDs, but it is still consistent with it. Because the RGB tip distance is based on a small number of stars, we adopt for Minni\,48 the distance derived from the position of the RC in the near-IR CMD.

\subsection{Metallicity and age determination}

Once the cluster reddening and distance are measured, a good fit to the PARSEC isochrones (Cassisi \& Salaris 1997, Bressan et al. 2012) can be secured, without much free space to vary the age and metallicity parameters.

The cluster metallicity can also be measured using the slope of the RGB in the near-IR CMD (e.g., Valenti et al. 2007, Cohen et al. 2017).
We measure a RGB slope of $S_{JK}=-0.11 \pm 0.05$ from the 2MASS decontaminated cluster CMD (Figure 7).
This slope is similar to that of the well studied metal-rich bulge GCs NGC\,6553 and NGC\,6528. 
Indeed, adopting the calibration of Cohen et al. (2017),  a metallicity $[Fe/H]=-0.2 \pm 0.3$ dex is derived, similar to those of the mentioned  two bulge GCs.  
The final metallicity $[Fe/H]=-0.2 \pm 0.3$ dex for Minni\,48 was confirmed through the fitting of PARSEC stellar isochrones.

The age of Minni\,48 is trickier to measure, as the Gaia photometry is not deep enough to reach the main-sequence turn-off (MSTO). 
The DECAPS photometry would be deep enough, but it is contaminated by the bright field stars. 
The near-IR photometry is also not deep enough to reach the cluster MSTO. 
However, a good fit to all the CMDs (Figures 5, 6, 7) is obtained using the PARSEC isochrones with $[Fe/H]=-0.2$ dex and an age of $10 \pm 2$ Gyr.
This error is determined by changing the isochrones at a fixed metallicity until they do not fit the cluster sequence in all CMDs.
We note that even though Minni\,48 is a bulge GC according to its position, metallicity and kinematics, 
its age appears to be somewhat younger than the ages of typical metal-rich GCs, which are $12-13$ Gyr old (e.g., Dotter et al. 2010).
Also, we note that even though  the bulge field stars are dominated by an old population, it also appears to exhibit a wide range of ages 
(e.g., Haywood wt al. 2016, Bensby et al. 2017, Bernard et al. 2018).
Obviously, our age measurement for this GC needs to be confirmed with deeper PSF photometry reaching well beyond the MSTO.

\subsection{Radius estimate}

Even though there is a bright foreground star contaminating the cluster field, Minni\,48 cluster looks to be quite bright in the near- and mid-IR images. 
Figure 13 shows the radial density profile constructed by performing star counts within $15'$ from the GC centre. Such profile was estimated using the centroid of the bright mid-IR stars.
The innermost bins have been ignored as they are clearly incomplete,{ in part,}  due to the presence of the bright saturated star IRAS 17302-2800 (clearly seen in Figures 2 and 3).

{ Given that the quality of the Gaia photometry may be substantially degraded in the crowded inner regions of GCs, especially those located in the Galactic Bulge, a significant fraction of stars in the inner region Minni\,48 present missing values of BP and/or RP bands. This also leads the radial density profile to dive below the stellar background level in its inner bin. This effect is absent in the cluster outskirts, as shown in Figure 13 and previously predicted by Riello et al. (2018). We argue that, except for the cluster core, the radial density profile is consistent with that of a GC, including the cluster radius of $6\pm1'$ suggested by the Figure 13.
}    

The angular size of this GC is estimated to be $r \sim 6' \pm 1' $, equivalent to $r \sim 14 \pm 2$ pc at the distance of the bulge,
a size that is comparable to those of other typical GCs. 
{ Note that this size should be taken as a rough estimate, given the difficulties in obtaining the surface density profile.}

\subsection{Measurement of total luminosity}

The total absolute magnitude (luminosity) of Minni\,48 is estimated to be $M_{Ks}= -9.04 \pm 0.66$ mag, adopting a distance of $D=8.4$ kpc.
This value was measured by coadding all bright cluster stars in the decontaminated near-IR CMD within 5' from the GC centre.
This is equivalent to $M_V=  -6.5 \pm 0.8$ mag, assuming a typical GC colour of $(V-K_s) \approx 2.5 \pm 0.5$ mag.
However, we stress that this estimated value is a lower limit, because of the contamination from the bright saturated star that does not allow us to count the central GC members.
 
Finally, we would also like to highlight the difficulties in the size determination if the bright star causes a shift in the determination of the centroid of the stellar density.

As a consistency check on the luminosity, this cluster appears to be quite bright in the GLIMPSE and WISE images.
In particular, the WISE image of Minni\,48 (Figure 3) resembles those of the known metal-rich bulge GCs, appearing somewhat brighter than NGC 6528 
($D=7.9$ kpc,  $M_V=-6.57$ mag), and also slightly fainter than NGC 6553 ($D=6.8$ kpc, $M_V=-7.77$ mag).

At first sight it may seem surprising that a moderately cluster like this was missed before. However, there are two other examples of similar GCs recently discovered at very low latitudes under similar circunstances: FSR 1758 (Cantat-Gaudin et al. 2018, Barb\'a et al. 2019), and VVV-CL160 (Borissova et al. 2014, Minniti et al. 2021a).
The first one is a much larger and brighter globular cluster ($Mi<-8.6$ mag, $r >15$’) that has been recently characterized using similar Gaia, VVV and DECAPs data (Barb\'a et al. 2019), and confirmed spectroscopically (Villanova et al. 2019).
The second example is a GC that has $M_K=-7.6$, equivalent to $M_V=-5.1$, and it is nearby (D=4.2 kpc) and extended ($R_t = 52$'), but it is invisible in the optical images, while being barely visible in the near-IR images. This GC was clearly seen thanks only to the Gaia EDR3 and VVV PMs that allowed us to discriminate the field population (Minniti et al. 2021a).

\subsection{Radial Velocities}

There are no radial velocities (RVs) available for obvious stellar sources associated with this GC. 
Unfortunately, most of its member stars are fainter than the limit of Gaia spectroscopy, which is $G\approx 15$ mag. 
Ground-based follow-up spectroscopic observations are certainly needed. 
Future Surveys like 4MOST (Chiappini et al. 2019, Bensby et al. 2019) would be able to easily reach the GC member giants, thus providing accurate RVs.

The final GC parameters adopted here are summarized in Table 1.

\section{The RR Lyrae Puzzle}

Minniti et al. (2018) singled out this cluster as a metal-poor GC candidate because of an apparent ($\sim 2 \sigma$) excess  of RR Lyrae stars in this field.
Table A1 lists all RR Lyrae within $8'$ from the GC, compiled 
from the OGLE (Soszy\'nski et al. 2014), VVV (Majaess et al. 2018) and Gaia (Clementini et al. 2019) catalogues.
The top panel of Figure 2 shows the location of RR Lyrae in the zoomed $9’ \times 7’$ $JHK_s$ image centred on Minni\,48.
The white arrows point to the RR Lyrae stars in the field and the yellow arrow to a Type 2 Cepheid (WVir type, Soszy\'nski et al. 2017).

Navarro et al. (2021) recently compiled all the bulge RR Lyrae variable stars from different Surveys in order to study their spatial distribution.
Figure 14 shows the distribution of distance moduli for all bulge RR Lyrae stars (grey histogram),
compared with the RR Lyrae located within $3'$ and $5'$ from the cluster centre (purple and green histograms, respectively). 
These distance moduli were computed following the procedure described by Minniti et al. (2017b) and Palma et al. (2019).
The black arrow indicates the adopted GC distance modulus  $m-M=14.61$ mag, for reference. 
As expected, the distance distribution for all RR Lyrae is centred
at a distance modulus of $14.5$ mag, being very broad (grey histogram) due to the depth of the bulge along the line of sight.
The  distances for the RR Lyrae stars  in the GC field are consistent with that of Minni\,48, although their Gaia EDR3 PMs are more scattered than those of the GC members 
as shown in Figure 14).
That can be due, however, to the low luminosity of these objects.
We also note, as a consistency argument, that adopting a much closer distance to these objects would lead to a very high reddening that would be inconsistent with the existing reddening maps.

On the basis of our confirmation of Minni\,48 as a metal-rich bulge cluster (based on the statistical decontamination of the optical and near-IR CMDs), 
these RR Lyrae stars are probably mere field objects and not cluster members, since RR Lyrae stars in metal-rich GCs are very rare.
The RR Lyrae appear to have a more extended distribution than that of the reddest giants that are clearly more clustered (Figure 9), and their PMs are also do not appear to be very clustered (Figure 15).
However, follow-up observations (especially measurements of RVs) of these objects are needed in order to confirm or discard cluster membership.
We tentatively conclude that these RR Lyrae are field objects because the GC is metal-rich, their PM distribution is not very concentrated, and they exhibit a wide distance distribution. However, we cannot definitely discard the presence of two stellar populations, nor that a couple of RR Lyrae may still be associated with the cluster.

Minni\,48 is certainly a very puzzling GC because in the optical Gaia CMD of this region there are two well separated RGBs (Figure 1). 
On one hand, the conspicuous RC and the redder GB seen in the near-IR data, added to the presence of very bright red stars in the WISE data, support a very metal-rich component.
On the other hand, the presence of RR Lyrae stars and the bluer GB suggest also the existence of a metal-poor population component.

The findings, however puzzling, are very interesting as the cluster region appears to contain a double population: a metal-rich old component with $[Fe/H]=-0.2$ dex, and a metal-poor old component with $[Fe/H] \sim -1.3$ dex.
If these results are confirmed spectroscopically not to be simply the effect of field stars contamination and the RR Lyrae stars are metal-poor cluster members, we may be either in the presence of the remaining nucleus of an accreted dwarf galaxy that originally had a composite population, or the final product of a merger between a metal-poor and a metal-rich cluster. 
In both regards, this is a very interesting object to be followed-up spectroscopically.

We also note that there is a planetary nebula/symbiotic star candidate (PN Th 3-30, Mikolajewska et al. 1997), located about 6' away from the cluster centre. 
This is also worth noticing because  planetary nebulae that are GC members are certainly very rare (Jacoby et al. 1997, Minniti et al. 2019).
Nevertheless, the association of this planetary nebula with Minni\,48 in this crowded field also needs to be confirmed spectroscopically.  

\section{Summary and  Conclusions}

We analyzed optical and near-IR photometry of the candidate GC Minni\,48, recently discovered using the VVV Survey. % (Minniti et al. 2018). 
The cluster field contains a bright foreground star, so multicolour photometric data were needed in order to confirm its real GC nature, as well as to determine its physical parameters.

We measured the cluster's fundamental parameters, including  reddening, extinction, distance, metallicity, age, radius, and luminosity.
Our main results for Minni\,48 are listed in Table 1. 
These parameters are  consistent with a luminous metal-rich GC, located at a projected angular distance of about 0.4 kpc from the Galactic centre.
{ 
We emphasize that the Minni\,48 analysis  took place by using  multiple photometry and a set of filters and decontamination procedures and  that the GC's evolutionary sequences survived in all procedures. It is unlikely that a field fluctuation survives such a set of well established tools.} 
Therefore, Minni\,48 joins the MW GC system as another genuine GC in the inner bulge.

{
Unfortunately, as a bulge GC, Minni\,48 accumulates a set of Gaia data limitations, such as bright stars in the projected neighbourhood and crowding effects within GCs core and Galactic Bulge, which significantly reduced the known gain provided by Gaia in star clusters analysis. Thus, deeper photometry is required to uncover the cluster core, but out of the core a prominent cluster can be seen. We also measure a cluster radius of $6\pm1'$, which is consistent with the typical sizes of Galactic GCs .
}

This is a very interesting bulge GC to follow up with more detailed observations.
In spite of the statistical decontaminated CMDs yielding only a metal-rich population, we cannot ignore the detection of a significant metal-poor stellar component in this field.
In particular, we also point out the presence of RR Lyrae variable stars in the field that need to be followed up spectroscopically in order to confirm or discard their cluster membership.

%%%%%%%%%%%%%%%%%%%%%%%%%%%%%%%%%%%%%%%%%%%%%%%%%%%%%%%%%%%%%%%%%%%%%%

\begin{acknowledgements}
We gratefully acknowledge data from the ESO Public Survey program ID 179.B-2002 and 198.B-2004 taken with the VISTA telescope, and products from the Cambridge Astronomical Survey Unit (CASU). 
This publication makes use of data products from the WISE satellite, which is a joint project of the University of California, Los Angeles, and the Jet Propulsion Laboratory/California Institute of Technology, funded by the National Aeronautics and Space Administration. This research has made use of NASA’s Astrophysics Data System Bibliographic Services and the SIMBAD database operated at CDS, Strasbourg, France, and the Two Micron All Sky Survey (2MASS), and the Gaia-DR2 and EDR3.
DM is supported by the BASAL Center for Astrophysics and Associated Technologies (CATA) AF B-170002, and by FONDECYT Regular grant No. 1170121.
J.A.-G. acknowledges support from Fondecyt Regular 1201490 and from ANID – Millennium Science Initiative Program – ICN12\_009 awarded to the Millennium Institute of Astrophysics MAS.
R.K.S. acknowledges support from CNPq/Brazil through project 305902/2019-9.
T.P. and J.J.C. acknowledge support from the Argentinian institutions CONICET and SECYT (Universidad Nacional de Córdoba). 

\end{acknowledgements}

\begin{figure}
\begin{centering}
\includegraphics[scale=0.47]{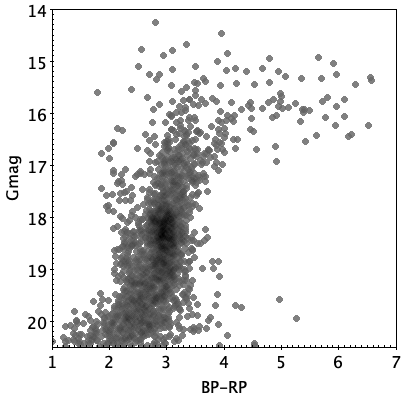}
\caption{
Gaia EDR3 optical CMD (G vs BP-RP) for a $r<3'$ region centred in the GC. 
This CMD shows that the upper RGB in this area is clearly split into a bluer and a redder component.
}
\label{fig:vvv_vhs}
\end{centering}
\end{figure}

\begin{figure}
\begin{centering}
\includegraphics[scale=0.38]{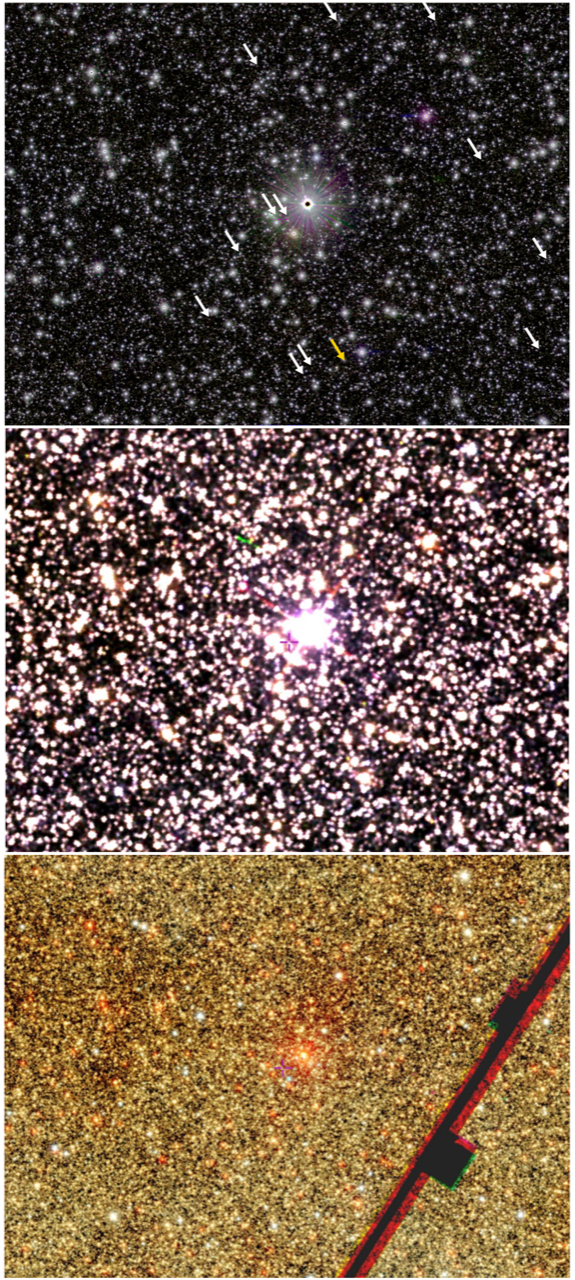}
\caption{Zoomed $9’ \times 7’$ region centred on Minni\,48 obtained with the VVV (top panel), 2MASS (middle panel), and DECAPS (bottom panel)
This figure  is oriented along equatorial coordinates, with East to the left and North to the top
 (see also Figure 3).
The white arrows point to the RR Lyrae stars and the yellow arrow to the Type 2 Cepheid.
The saturated star near the centre is the bright source IRAS 17302-2800 with $K_s=3.33$ and $(J-K_s)=1.81$ mag. 
}
\label{fig:vvv_vhs}
\end{centering}
\end{figure}

\begin{figure}
\begin{centering}
\includegraphics[scale=0.35]{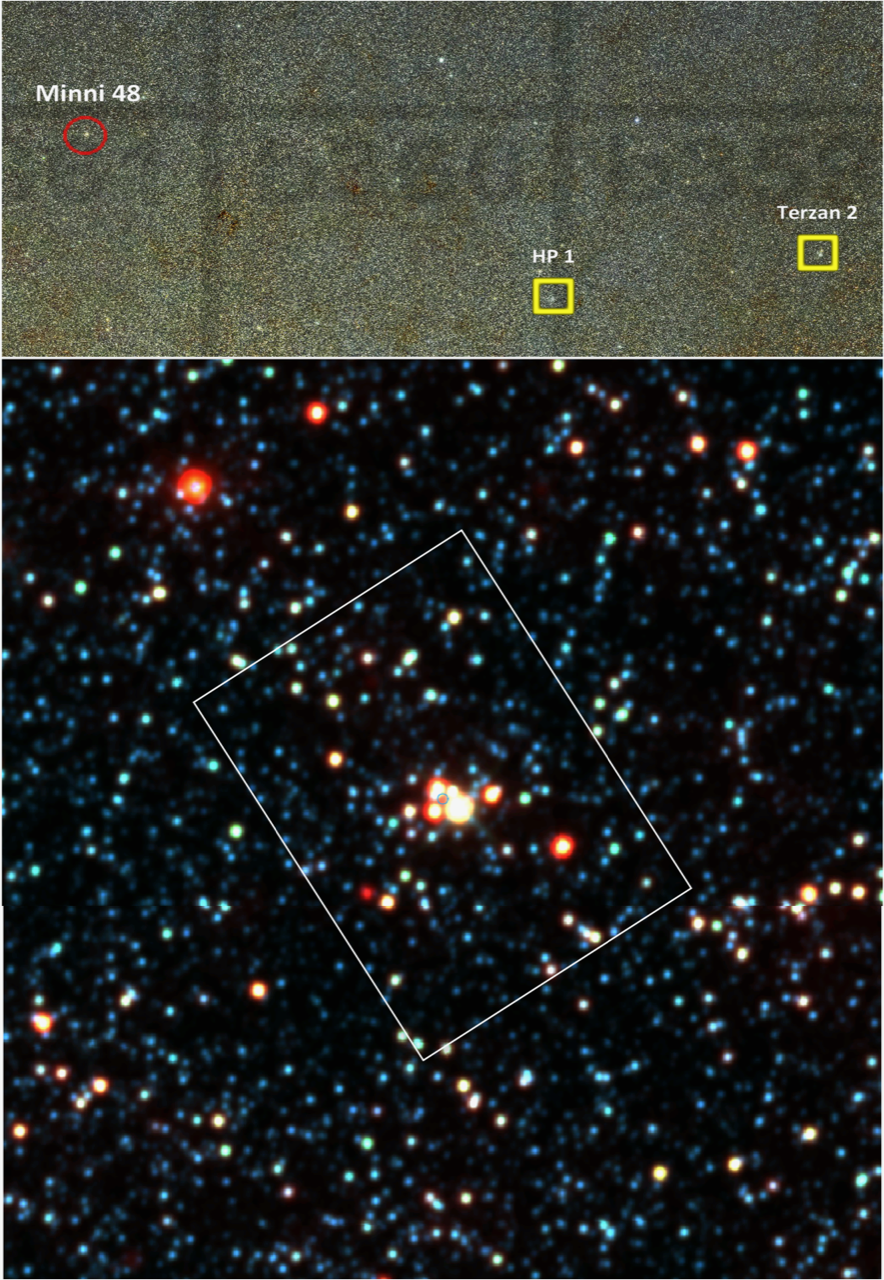}
\caption{
Top: Wide field $JHKs$ mosaic covering the VVV tiles b359 to b361, showing the positions of Minni\,48 compared with those of the two well known GCs: HP 1 and Terzan 2.
This field covering approximately $1.5^{\circ} \times 4^{\circ}$ is oriented along Galactic coordinates, with longitudes increasing to the left and latitudes to the top.
Bottom: Zoomed $20’ \times 20’$ region centred on Minni\,48 obtained with WISE in the mid-IR (3.6 to 8 microns).
There is a clear concentration of bright IR stars in the cluster core. The white box shows the location of the zoomed images portrayed in Figure 2.
}
\label{fig:vvv_vhs}
\end{centering}
\end{figure}

\begin{figure}
\begin{centering}
\includegraphics[scale=0.30]{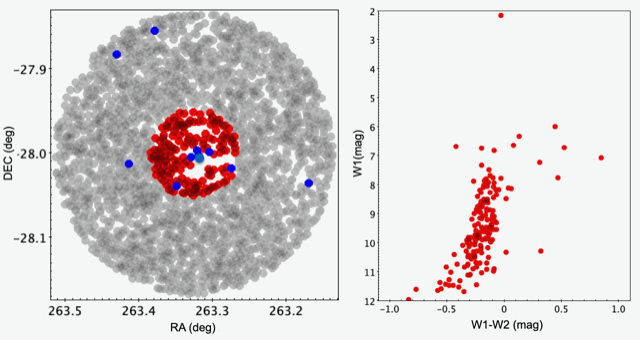}
\caption{
Map of WISE sources (left panel) and WISE mid-IR CMD (right panel). There is a clear concentration of many RGs in the cluster core. 
The brightest IR stars are indicated with blue circles.
The brightest star with $W1=2.1$, $W1-W2=0$ mag is the foreground source IRAS 17302-2800 located close to the cluster centre.
For comparison, the region where the Gaia-EDR3 CMD of Figure 1 is made ($r<3'$) is the inner red region shown in this map.
}
\label{fig:vvv_vhs}
\end{centering}
\end{figure}

\begin{figure}
\begin{centering}
\includegraphics[scale=0.55]{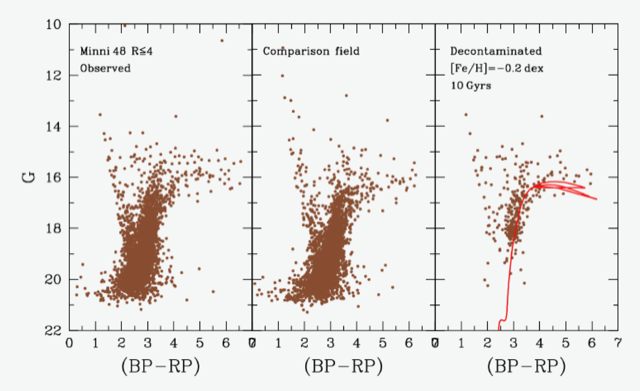}
\caption{
Gaia EDR3 optical CMDs (G vs BP-RP) for the GC region (left panel), a background region (middle panel), and the statistically decontaminated region following Camargo et al. (2009) (right panel). 
The latter  clearly shows that only the  RGB (more metal-rich) survives the decontamination procedure.
The Gaia decontaminated  CMD is fitted by a PARSEC isochrone with age $10$ Gyr and metallicity $[Fe/H] = -0.2$ dex.
}
\label{fig:vvv_vhs}
\end{centering}
\end{figure}

\begin{figure}
\begin{centering}
\includegraphics[scale=0.25]{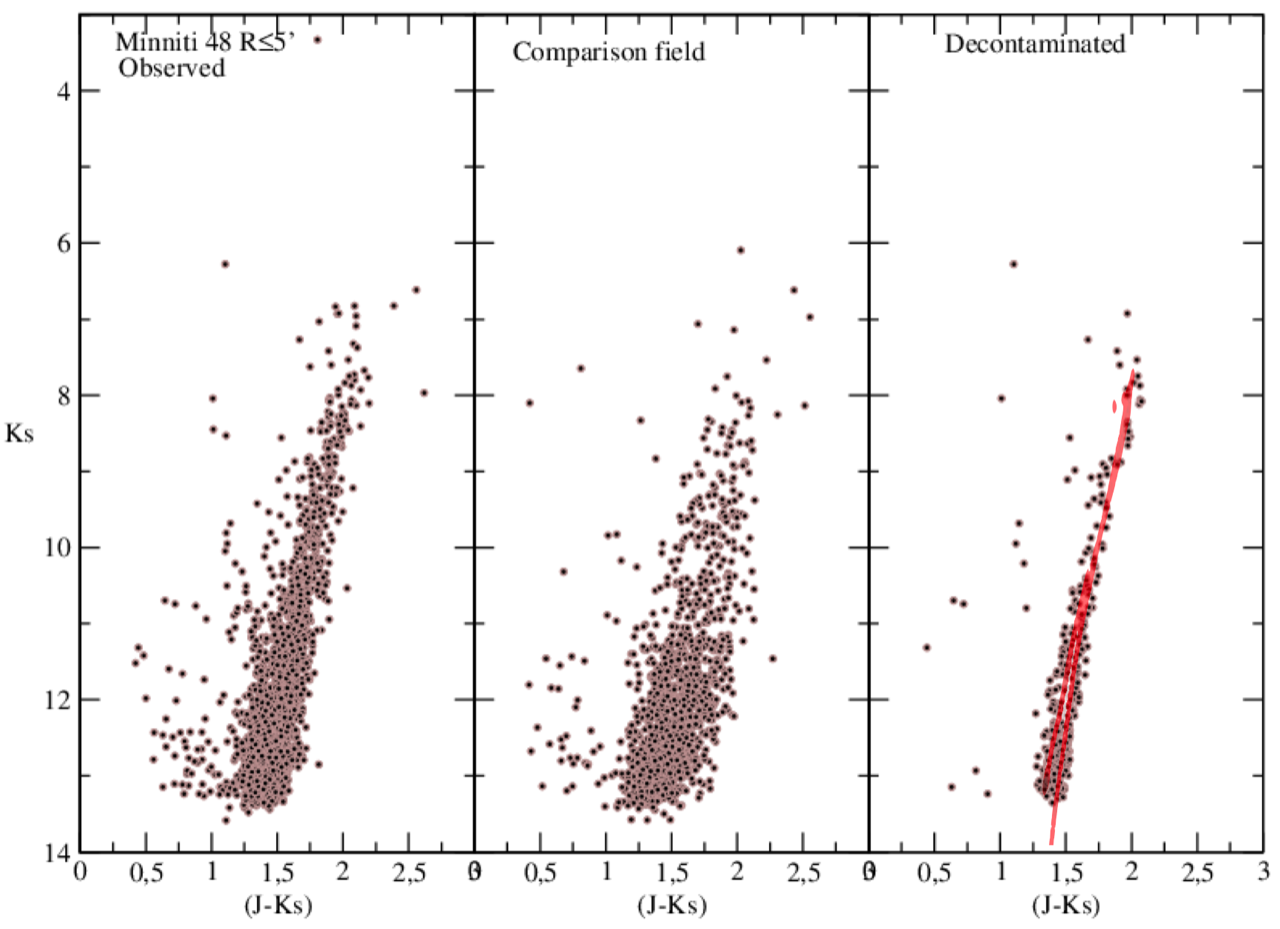}
\caption{
2MASS near-IR CMDs ($K_s$ $vs$ $J-K_s$) for the GC region (left panel), a background region (middle panel), and the statistically decontaminated region following Camargo et al. (2009) (right panel).  The latter clearly shows a tight metal-rich RGB and the RGB tip at $K_s=7.75 \pm 0.10$ mag. 
The red line shows a PARSEC isochrone of age $10$ Gyr, and metallicity $[Fe/H] = -0.2$ dex.
}
\label{fig:vvv_vhs}
\end{centering}
\end{figure}
\begin{figure}
\begin{centering}
\includegraphics[scale=0.3]{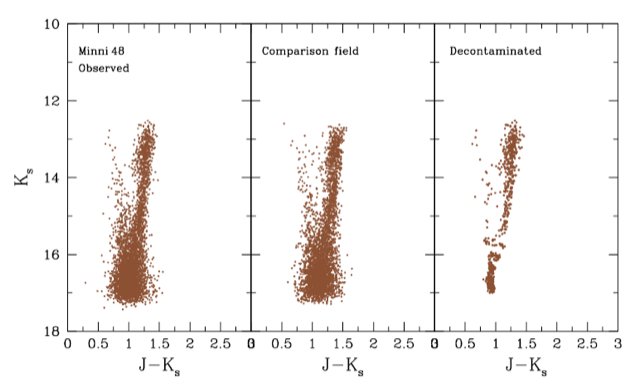}
\caption{
Deep VVV near-IR CMDs ($K_s$ vs $J-K_s$) for the GC region (left panel), a background region (middle panel), and the statistically decontaminated region following Palma et al. (2018) (right panel). 
The latter clearly shows the GC RC at $K_s=13.45$ mag. 
}
\label{fig:vvv_vhs}
\end{centering}
\end{figure}

\begin{figure}
\begin{centering}
\includegraphics[scale=0.3]{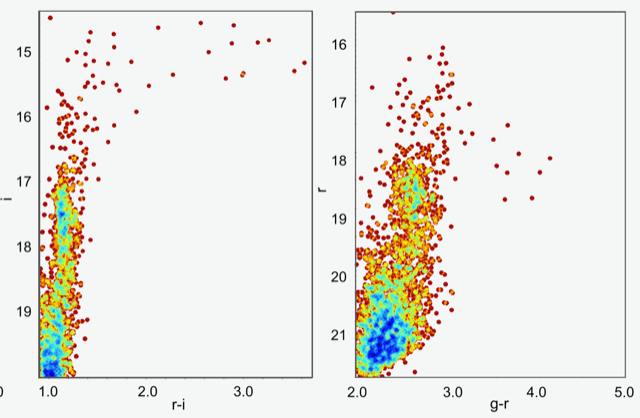}
\caption{
Optical CMDs within 3' from the GC center using the DECAPS Survey (Schlafly et al. 2017). 
These diagrams clearly show the GC RC at $r=18.65$ and $i=17.50$ mag. 
}
\label{fig:vvv_vhs}
\end{centering}
\end{figure}

\begin{figure}
\begin{centering}
\includegraphics[scale=0.37]{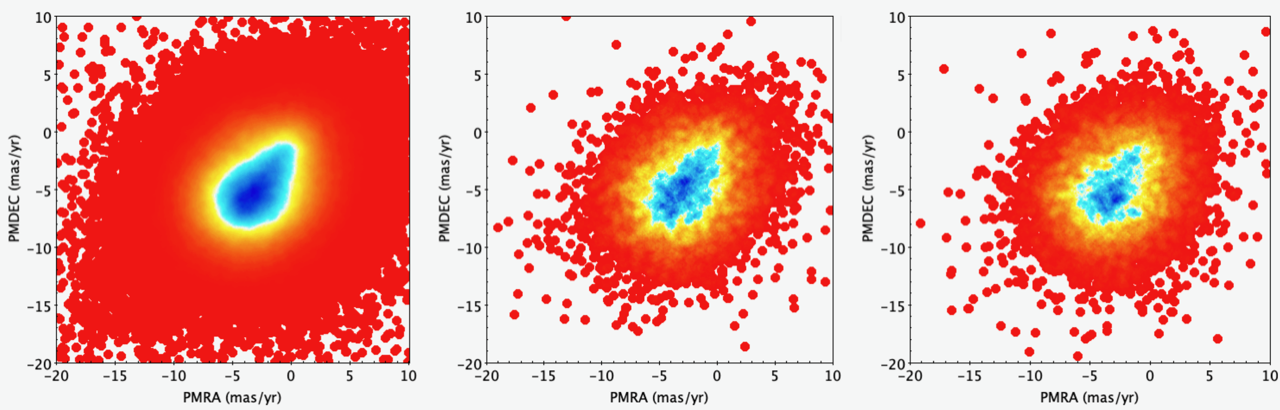}
\caption{
Gaia-EDR3 vector PM diagrams for the whole 20' field (left panel) compared with a 5' field centred on the cluster (right panel) and with an equal area field located 10' South of the cluster (middle panel
).
}
\label{fig:vvv_vhs}
\end{centering}
\end{figure}

\begin{figure}
\begin{centering}
\includegraphics[scale=0.2]{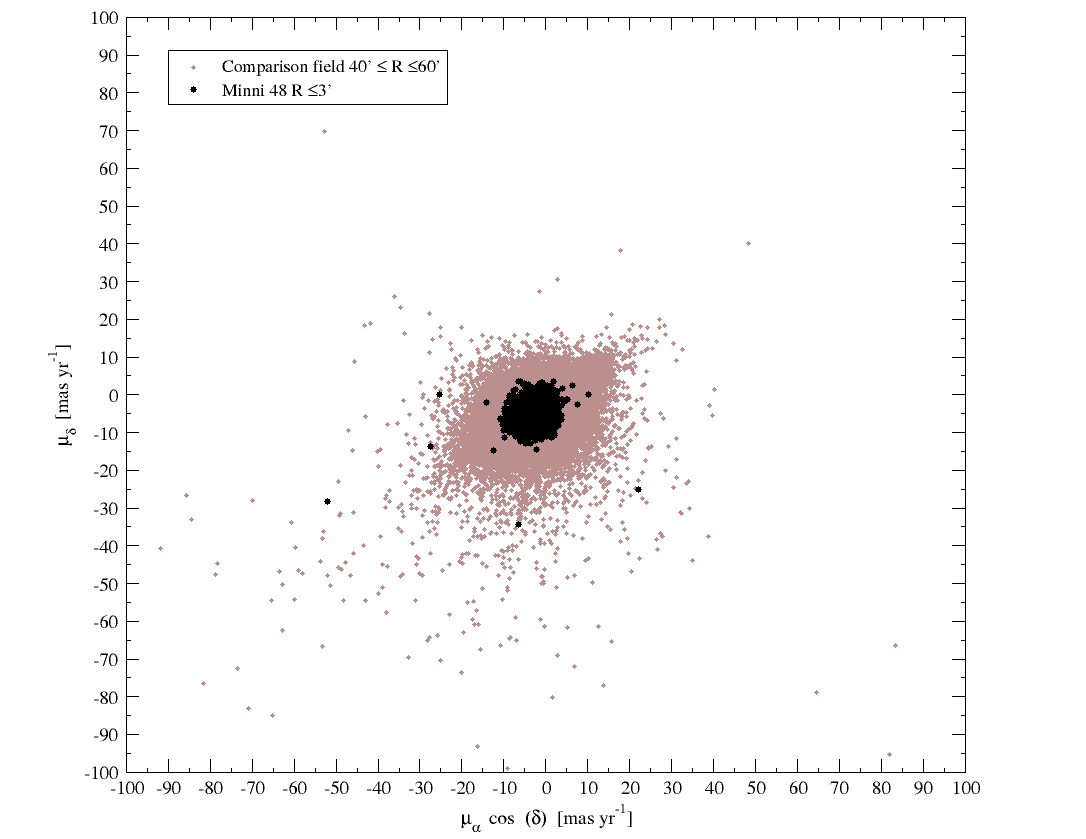}
\caption{Gaia-EDR3 PM distribution. The black circles are the stars in the central region of Minni\,48 (errors $<0.3\,mas/yr$), while the brown circles represent
stars located in a ring in the surrounding field.}
\label{fig:vvv_vhs}
\end{centering}
\end{figure}

\begin{figure}
\begin{centering}
\includegraphics[scale=0.22]{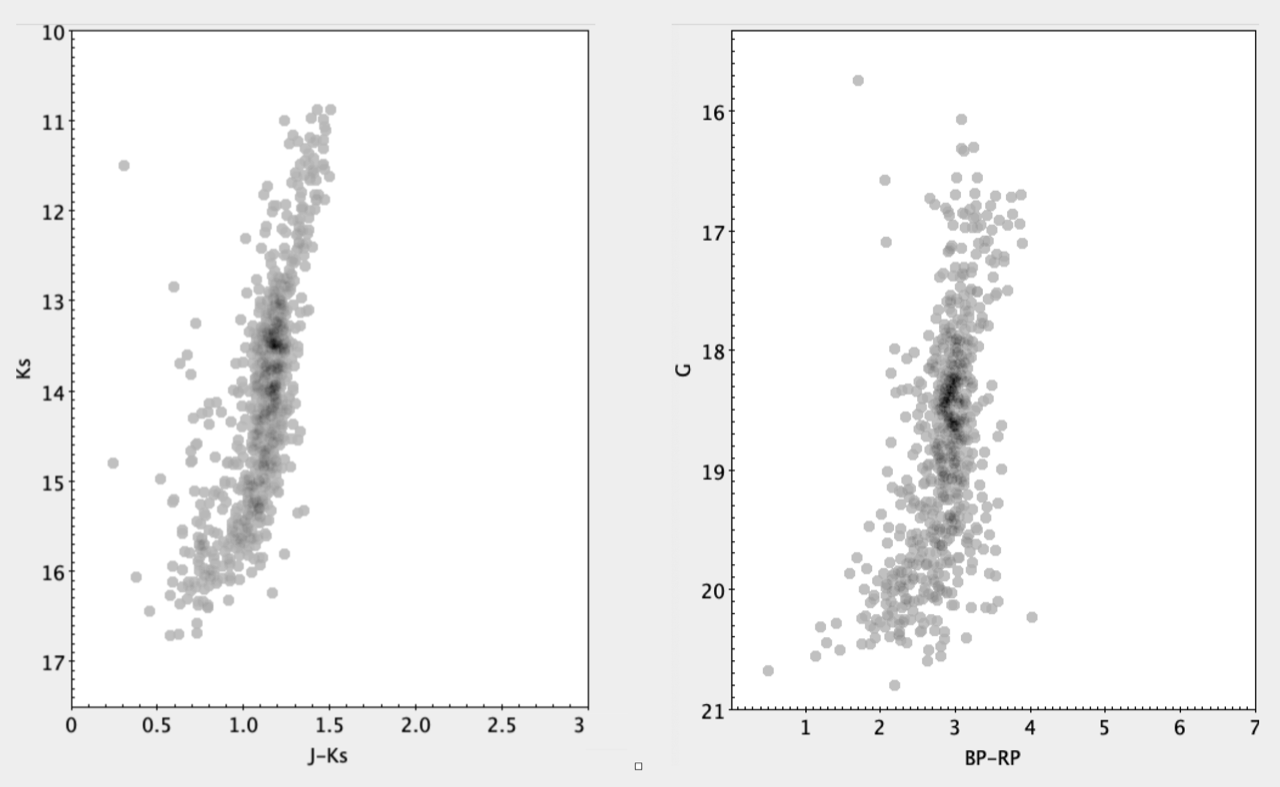}
\caption{
Optical and near-IR CMDs for the GC region after the PM decontamination using Gaia EDR3 data.
}
\label{fig:vvv_vhs}
\end{centering}
\end{figure}

\begin{figure}
\begin{centering}
\includegraphics[scale=0.4]{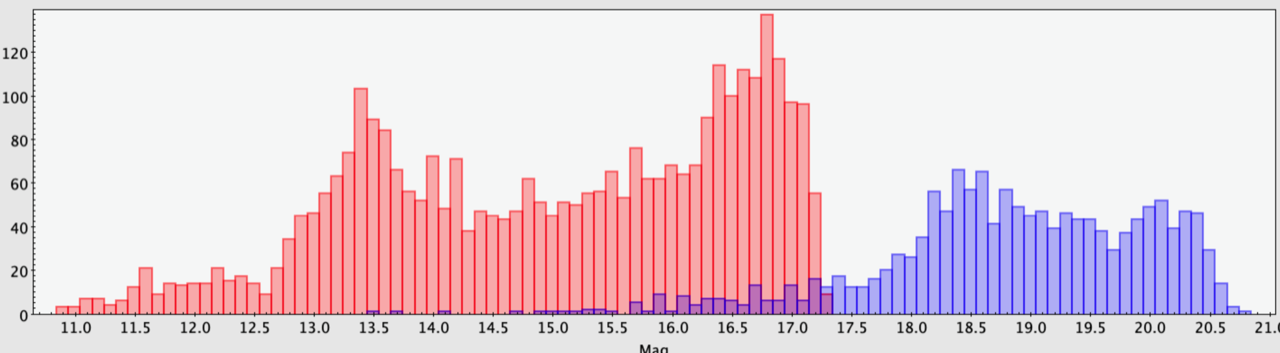}
\caption{
Optical and near-IR CMDs luminosity functions (in blue and red, respectively) for the GC region after the PM decontamination using Gaia EDR3 data.
}
\label{fig:vvv_vhs}
\end{centering}
\end{figure}

\begin{figure}
\begin{centering}
\includegraphics[scale=0.23]{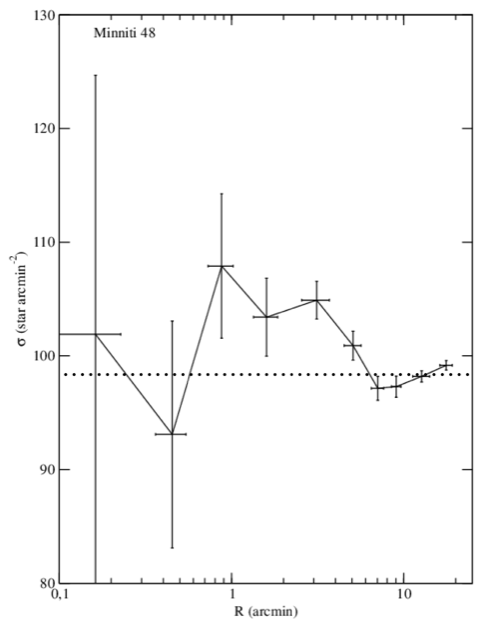}
\caption{
Stellar surface density as function of distance from the cluster centre. 
Note that the star counts in the innermost bins are affected by the bright saturated star. 
The dotted horizontal line indicates the adopted value for the background.
We measure a cluster radius $r=6' \pm 1'$.
}
\label{fig:vvv_vhs}
\end{centering}
\end{figure}

\begin{figure}
\begin{centering}
\includegraphics[scale=0.36]{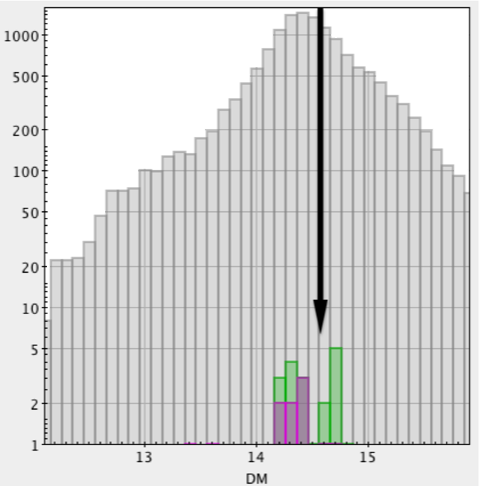}
\caption{
Distribution of distance moduli for all bulge RR Lyrae stars from the OGLE, VVV and Gaia catalogues (grey histogram), compared with the RR Lyrae stars located within 3' and 5' from the cluster centre (purple and green histograms, respectively). 
The black arrow indicates the adopted GC distance modulus for reference.
}
\label{fig:vvv_vhs}
\end{centering}
\end{figure}

\begin{figure}
\begin{centering}
\includegraphics[scale=0.28]{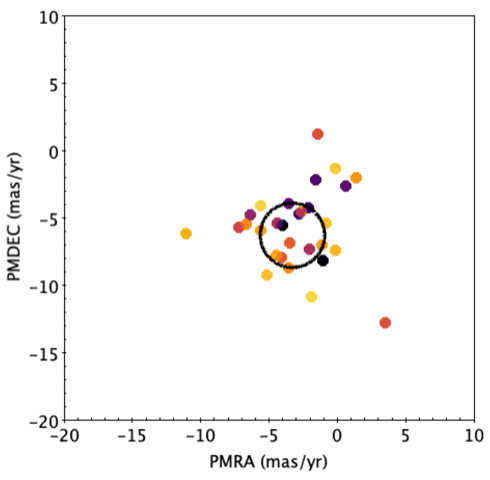}
\caption{ Vector PM diagram for all bulge RR Lyrae stars from the OGLE, VVV and Gaia catalogues  located within 10' from the cluster centre. The scale is the same as in Figure 9, and
the RR Lyrae are color-coded with distance from the GC centre, with the  darkest objects being closer to the centre, and the lighter ones being farther away.
The black circle is centred on the mean GC PM for comparison.}
\label{fig:vvv_vhs}
\end{centering}
\end{figure}

\begin{table*}
\caption[]{Summary of the derived physical parameters for Minni\,48}
\label{tab:strategy}
\begin{center}
\begin{tabular}{lccll}
\hline
Parameter & Value  &  &  \\
\hline
RA(J2000)  & $17h33m18.0s$ &  &  \\
DEC(J2000) & $-28d00m02s$ &  &  \\
Longitude & $359.3513$ deg&  &  \\
Latitude &   $2.7903$ deg&  &  \\
Radius & $6' \pm 1'$  &  &  \\ 
G (Red Clump)  & $16.10 \pm 0.10$ mag &  &  \\
$K_s$ (Red Clump) & $13.30 \pm 0.10$ mag &  &  \\
$A_G$               & $3.23 \pm 0.10$ mag &  &  \\
$A_{Ks}$           & $0.45 \pm 0.05$ mag  &  &  \\ %(Schlafly 0.172)
$E(J-K_s)$          & $0.61 \pm 0.03$ mag  &  &  \\ %(Gonzalez mide 0.21 mag)(Schlafly mide 0.23 mag)
Distance            & $8.4 \pm 0.3$ kpc &  &  \\
$R_{gal}$          & $1.09$ kpc &  &  \\
$PM_{RA}$       & $-3.5 \pm 0.5$ mas/yr &  &  \\
$PM_{DEC}$    & $-6.0 \pm 0.5$ mas/yr &  &  \\
Abs. Magnitude $M_{Ks}$ & $ -9.04 \pm 0.66$ mag &  &  \\
$[Fe/H]$            & $-0.20 \pm 0.3$ dex &  &  \\
Age                   & $10 \pm 2$ Gyr &  &  \\
\hline
\end{tabular}
\end{center}
\end{table*}

\begin{appendix}

{
Table A1 below lists the data for all the known RR Lyrae variable stars in the cluster field (within 8'). 
We give the IDs, periods in days, distances from the cluster centre in arcminutes,
positions in equatorial coordinates (J2000), optical and near-IR magnitudes from OGLE and VVV, respectively, and RR Lyrae types. 
}

\begin{table*}
\begin{center}
\caption[]{Photometry of the RR Lyrae variables in the cluster field (within 8')}
\begin{tabular}{lccccccccc}
\hline
ID &  Period (d) & Dist(')   & RA(J2000) & DEC(J2000) & V & I & J & Ks & Type \\
\hline
OGLE-BLG-RRLYR-20594 &0.37874882  &  2    &17 33 18.14 &-28 00 03.4     &19.105 &16.770  &14.874 &13.806  &RRc\\    %G=18.0740  PM   0.8 .3,6
OGLE-BLG-RRLYR-20598 &0.65394063  & 10    &17 33 18.58 &-27 59 55.4    &19.410 &17.015  &14.808 &13.674  &RRab\\  %G=18.3886  PM -4.0 -4.1  % gaia 4061164605179191040 P_gaia=0.40406 
OGLE-BLG-RRLYR-20640 &0.67065639  & 62    &17 33 22.53 &-27 59 45.5    &19.387 &16.901  &14.793 &13.996  &RRab\\  %G=18.1585  PM -1.3 -1.9
OGLE-BLG-RRLYR-20695 &0.30129839 &131    &17 33 27.90 &-27 59 54.0    &20.454 &17.786  &15.796 &14.936  &RRc\\    %G=19.2115   PM  0.1 -2.4
VVV-173309.6-275815      &0.5497144     &155    &17 33 09.58 &-27 58 14.9    & --       & --          &15.425 &14.524  &RRab \\  %G=
OGLE-BLG-T2CEP-438    &15.117615     &164    &17 33 25.71 &-28 02 09.5  & 16.270 & 13.734  &12.066 &10.096  &T2C  \\  %G=
OGLE-BLG-RRLYR-20680 &0.27799538 &157    &17 33 26.75 &-28 01 48.6    &19.335 &17.171  &15.474 &14.827  &RRc *\\  %G=18.4116  PM -0.8 -8.16% gaia 4061163707507695488 P_gaia = 0.27800
OGLE-BLG-RRLYR-20691 &0.56305356  &166    &17 33 27.68 &-28 01 47.0    &19.341 &17.118  &15.474 &14.827 &RRab *\\ %G=18.387   PM -6.0 -3.5
OGLE-BLG-RRLYR-20457 &0.56840434  &205    &17 33 03.31 &-27 58 57.9    &18.725 &16.309  &14.688 &13.607  &RRab\\ %G=17.6809  PM  -2.1 -0.4 gaia 4061176562379121280
OGLE-BLG-RRLYR-20485 &0.53964256  &208    &17 33 06.23 &-28 02 20.0    &19.816 &17.277  &15.150 &14.353  &RRab\\ %G=18.5671  PM  -2.3 -7.1
OGLE-BLG-RRLYR-20419 &0.49726189  &249    &17 32 59.24 &-28 00 22.8    &20.280 &17.754  &15.762 &14.819  &RRab\\ %G= 19.1347  PM -2.8 -5.5
OGLE-BLG-RRLYR-20752 &0.21171665  &271    &17 33 32.80 &-28 03 09.0    &20.878 &18.593  &16.414 &15.536  &RRc\\   %G=19.9358
OGLE-BLG-RRLYR-20519 &0.58090376  &272    &17 33 10.10 &-28 04 13.3    &19.641 &17.240  &15.231 &14.444  &RRab\\ %G=18.5841   PM -4.5 -5.3
OGLE-BLG-RRLYR-20578 &0.47401305  &294    &17 33 16.21 &-28 04 55.4    &20.333 &17.816  &16.155 &15.072  &RRab\\ %G=19.2285   PM -7.1 -5.1
OGLE-BLG-RRLYR-20353 &0.59896727  &337    &17 32 53.40 &-28 01 26.6    &20.320 &17.583  &15.521 &14.595  &RRab\\ %G=18.9952   PM -5.6 -4.9
VVV-173335.5-275512   &0.6389482  &371    &17 33 35.49 &-27 55 12.6          & --       & --           &15.761 &14.708  &RRab \\ %G=
VVV-173343.4-275608   &0.5289875  &410    &17 33 43.39 &-27 56 07.8           & --       & --          &15.527 &14.808  &RRab\\  %G=
OGLE-BLG-RRLYR-20297 &0.66086054  &413    &17 32 46.88 &-28 00 12.1    &20.013 &17.327  &15.199 &14.177  &RRab\\ %G=18.7323   PM -1.9 -10.2
VVV-173348.8-275820   &0.4748277  &420    &17 33 48.80 &-27 58 19.9          & --       & --           &15.826 &15.004  &RRab \\ %G=   Gaia 4061169999664021504 P= 0.51893
OGLE-BLG-RRLYR-20368 &0.57258412  &442    &17 32 54.96 &-28 04 55.4    &20.170 &17.585  &15.526 &14.708  &RRab\\ %G=18.9313   PM  -4.1 -5.4
VVV-173349.3-275706   &0.4895341  &451    &17 33 49.34 &-27 57 06.2          & --       & --           &15.689 &14.783  &RRab\\  %G=

\hline
\end{tabular}
\end{center}
\end{table*}

\end{appendix}

\end{document}